\documentclass[
10pt
,aps
,english
,longbibliography
,nofootinbib
,prx
,superscriptaddress
,twocolumn
]{revtex4-2}

\usepackage[english]{babel}
\usepackage{booktabs}
\usepackage{braket}
\usepackage{comment}
\usepackage{csquotes}
\usepackage{enumitem}
\usepackage{float}
\usepackage[T1]{fontenc}
\usepackage{graphicx} 
\usepackage[colorlinks
,breaklinks
,urlcolor={blue}
,linkcolor={red}
,citecolor={blue}]{hyperref}
\usepackage{lmodern}
\usepackage{mathrsfs}
\usepackage{mathtools}
\usepackage{pgfplotstable}
\usepackage{siunitx}
\usepackage{xurl}

\DeclareQuoteStyle{english}%
    {\textquoteleft}
    [\textquoteleft]
    {\textquoteright}
        [0.05em]   
    {\textquotedblleft}
    [\textquotedblleft]
    {\textquotedblright}

\graphicspath{ {figures/} }
\pgfplotsset{compat=1.18}
\begin{document}
\setquotestyle{english}
\title{Design and efficiency in graph state computation}
\begin{abstract}
The algorithm-specific graph and circuit etching are two strategies for compiling a graph state to implement quantum computation.  Benchmark testing exposed limitations to the proto-compiler, Jabalizer giving rise to Etch (\url{https://github.com/QSI-BAQS/Etch}), an open-source, circuit-etching tool for transpiling a quantum circuit to a graph state.  The viability of circuit etching is evaluated, both as a resource allocation strategy for distilling magic states and as an alternative to the algorithm-specific graph strategy as realised in Jabalizer.  Experiments using Etch to transpile IQP circuits to an equivalent graph state resulted in higher ratios of Pauli qubits to non-Pauli qubit than required for efficient magic state distillation.  Future research directions for the algorithm-specific graph and circuit-etching strategies are proposed.
\end{abstract}
\author{Greg Bowen}
\affiliation{Centre for Quantum Software and Information, University of Technology Sydney, Sydney, NSW 2007, Australia}
\author{Athena Caesura}
\affiliation{Zapata Computing, Boston MA 02110, USA}
\author{Simon Devitt}
\affiliation{Centre for Quantum Software and Information, University of Technology Sydney, Sydney, NSW 2007, Australia}
\author{Madhav Krishnan Vijayan}
\affiliation{Centre for Quantum Software and Information, University of Technology Sydney, Sydney, NSW 2007, Australia}
\maketitle

\section{Introduction}
The measurement-based quantum computing model (henceforth, MQBC) \cite{raussendorfOneWayQuantumComputer2001,raussendorfComputationalModelUnderlying2002,raussendorfMeasurementbasedQuantumComputation2003,briegelMeasurementbasedQuantumComputation2009} is an alternative to the circuit-based approach that is widely viewed as the \textit{de facto} standard for quantum computation.  Circuit-based computation proceeds by one or more unitary transformations (\enquote{gates}) of its component qubits whereas MQBC works through single-qubit measurements of a cluster of entangled qubits, drawing on the property of quantum teleportation to propagate state through the cluster.  Rau{\ss}endorf and Briegel \cite{raussendorfComputationalModelUnderlying2002} denote the cluster, $QC_{\mathcal{C}}$.  A graph, $G$ is an ordered pair of elements $V\{G\}$, a non-empty, finite set of vertices; and $E\{G\}$, a finite set of edges.  As a graph \textit{state}, $\ket{G}$, vertices are qubits and edges are interactions within a simple graph thereby prohibiting, 
\begin{itemize}
   \item more than one edge between vertices; and
   \item a vertex \enquote{self-joining} through an edge.
\end{itemize}
Note, $\ket{G}$ is configured as a two-dimensional lattice.  Expressing $QC_\mathcal{C}$ as $\ket{G}$ with qubits $(v\in V)$ in state $\ket{+}$ and $CZ$ interactions between qubits $((u,v)\in E)$,
\[
   \ket{G} = \prod_{(u,v)\in E\{G\}} CZ^{(u,v)} \left(\underset{v\in V\{G\}}\bigotimes \ket{+}_v\right).
   \label{eq1}
\]

A graph-state is also a stabiliser state.  An $n$ qubits stabiliser state, $\ket{\phi}$ is a simultaneous eigenvector with eigenvalue $+1$ of $n$ commuting and independent elements of the Pauli group, $\mathcal{P}^N$
\[
\label{PauliGroup}
\mathcal{P}^N\coloneqq\{\pm{\textit{I}},\pm{\textit{iI}},\pm\textit{X},\pm\textit{iX},\pm\textit{Y},\pm\textit{iY},\pm\textit{Z},\pm\textit{iZ}\}^N.
\]
The stabilisers of $\ket{\phi}$ are defined as,
\[
   \mathcal{S}\coloneqq\{M\in \mathcal{P}^N | M\ket{\phi} = \ket{\phi}\},
\]
being a group of operators with real overall phase $+1$ and $M$ is a generator of $\mathcal{S}$.  By extension, every vertex, $i$ of a graph state, $\ket{G}$ has a stabiliser, $K^{(i)}$ such that,
\[
\label{Ka}
   K^{(i)}=\sigma_x^{(i)} \bigotimes_{j\in E} \sigma_z^{(j)},
\]
where $E$ denotes the set of all vertices neighbouring $i$; thus, $\mathcal{S}$ for $G$ is the set, $M$ where,
\[
   \{M\} = K^{(i)}, i=(1 \dots N).
\]

\vspace{3mm}
This paper is a comparison of qubit resource allocation strategies as effected by the proto-compiler application, Jabalizer and the transpiler application, Etch.  Both applications, 
\begin{itemize}
   \item specify graph states to be passed to (hypothetical) quantum hardware for processing,
   \item draw on a universal quantum gate set comprising of Clifford group operators and the non-Clifford, $\frac{\pi}{4}$ ($T$) rotation \cite{barencoElementaryGatesQuantumComputation1995}.
\end{itemize}
Jabalizer \cite{vijayanCompilationAlgorithmSpecificGraphStates2022}, splits an input circuit between  Clifford operations, which it passes to a classical computer, and non-Clifford operations, which go to the quantum hardware.  In comparison, circuit etching, as defined by Etch, would pass all operations of the input circuit to the quantum hardware.  This paper is an evaluation of each application's method of resourcing the magic state distillation required to process the $T$ gate component of an input algorithm.  Section \ref{GSA} is background on Jabalizer and Etch as working prototypes of resourcing tools and the limitations to Jabalizer that gave rise to Etch.  Section \ref{evetch} is an outline of tests applied to Etch and the consistent results of a higher Pauli to non-Pauli qubits ratio than sought.  Section \ref{Disc} is analysis of section \ref{evetch} data, to establish whether Etch is a suitable component of a quantum compiler toolchain, both in its own right and as an alternative to Jabalizer.  The paper closes with suggestions for future research.

Note, the following nomenclature applies to this paper, 
\begin{itemize}
   \item Unless explicitly stated to the contrary, the term \enquote{qubit} refers to a \textit{logical} qubit.  A logical qubit is an elementary component of error-correction strategies that are necessary to fault-tolerant quantum computation.\footnote{ See \cite{gottesmanQuantumErrorCorrectionFaultTolerantQuantumComputation2009,devittQuantumErrorCorrectionForBeginners2009,fowlerSurfaceCodesTowardsPracticalLargeScaleQuantumComputation2012,terhalQuantumErrorCorrectionQuantumMemories2015} as entry points to error-correction strategies that are relevant to this paper.}
   \item The non-Clifford $T$ gate, a rotation of $\frac{\pi}{4}$ radians on the $Z$-axis, cannot be implemented directly in a fault-tolerant manner for error-corrected systems.  Processing a $T$ operation at an acceptable threshold of error is achieved through the process of magic state distillation \cite{bravyiUniversalQuantumComputationIdealCliffordGatesNoisyAncillas2005}, which also increases the overall number of qubit resources required to resolve a computation.  The term \textit{distillation ratio} will refer to the proportion of qubits recycled from a cluster state to the number of qubits required to effect the magic state distillation necessary to the computation.  The ratio itself is a target for Etch either to meet or approximate within an arbitrary level of tolerance, as a metric of resource allocation efficiency.
\end{itemize}

\section{graph-state architectures}\label{GSA}
The graph state has received considerable attention as a modelling framework \cite{schlingemannLogicalNetworkImplementationClusterStatesGraphCodes2002,schlingemannClusterStatesAlgorithmsGraphs2004,vandennestGraphicalDescriptionActionLocalCliffordTransformationsGraphStates2004,heinEntanglementGraphStatesApplications2006,andersFastSimulationStabilizerCircuitsUsingGraphState2006,dahlbergHowTransformGraphStatesUsingSingleQubitOperations2020} (also, \cite{nielsenClusterstateQuantumComputation2006,browneOneWayQuantumComputation2006}); as the name suggests, a graph state is a quantum state configured as a (simple) graph, $G$.  In the case of $QC_\mathcal{C}$, computation proceeds by measuring single qubits in a certain order and basis with overall quantum information propagating left-to-right.  Each Clifford group operator - Hadamard, $S$, CNOT - is a prescribed combination of measurements from $\mathcal{P}^N$ \cite{raussendorfComputationalModelUnderlying2002} and these measurements can occur simultaneously.  After \cite{danosExtendedMeasurementCalculus2009}, each element of the universal quantum gate set as used by Jabalizer/Etch will be termed a \textit{pattern}.  Further, patterns are tractable and their output can be linked through composition to form more complex widgets \cite{raussendorfMeasurementbasedQuantumComputation2003, danosExtendedMeasurementCalculus2009}.  Irrespective of the stochastic outcome of measuring qubit $j$ of $QC_\mathcal{C}$, MBQC is a deterministic computing model.  Each measurement of $QC_\mathcal{C}$ creates a byproduct, which in turn determines the method of measuring a readout qubit.  If
\begin{itemize}
   \item readout measurements are in a Pauli basis, they are \enquote{interpreted} according to their byproduct; otherwise,
   \item previous measurements determine the basis of readout measurement. 
\end{itemize}
See \cite{raussendorfComputationalModelUnderlying2002} for a fuller description of byproducts and contingent measurement of readout qubits.

Transforming $\ket{G}$ by means of a $T$ operation imposes an extra resourcing cost of the noiseless qubits needed for magic state distillation.  Briefly, the required $T$ state is distilled by preparing a number of ancillary magic state qubits with an equal number of noiseless qubits to encode one qubit at a low probability of error.  This same process as applied to $\ket{G}$ must result in a single low-error encoded readout qubit that will input to any subsequent pattern.  The ancillary qubits required to process a $T$ pattern in $QC_\mathcal{C}$ will constitute their own lattice, $\ket{G}_T$, separate from $QC_\mathcal{C}$ and henceforth termed a $T$ \textit{factory}.  As the initial step, it takes 15 physical qubits to encode one logical qubit in a $T$-state \cite{litinskiGameSurfaceCodes2019,litinskiMagicStateDistillation2019}.  This 15-to-1 magic state distillation encodes the $T$-state at probability of error, 35\textit{p}\textsuperscript{3} to leading order.  As might be expected, it is possible to realise magic states of a higher fidelity at the cost of more qubits.  The important consideration is that each instance of a $T$ operator in an algorithm necessarily increases the qubit requirements of the $QC_\mathcal{C}$ fulfilling that algorithm  \cite{raussendorfTopologicalFaultToleranceClusterStateQuantumComputation2007}.  Irrespective of the desired probability of error, all qubits necessary to processing the input algorithm must feature in $QC_\mathcal{C}$ at compile time or otherwise be obtained \textit{at some stage during runtime}.  It remains now to introduce the applications, Jabalizer and Etch, and their respective resourcing estimations.

\begin{figure*}
   \centering
   \includegraphics[width=\linewidth]{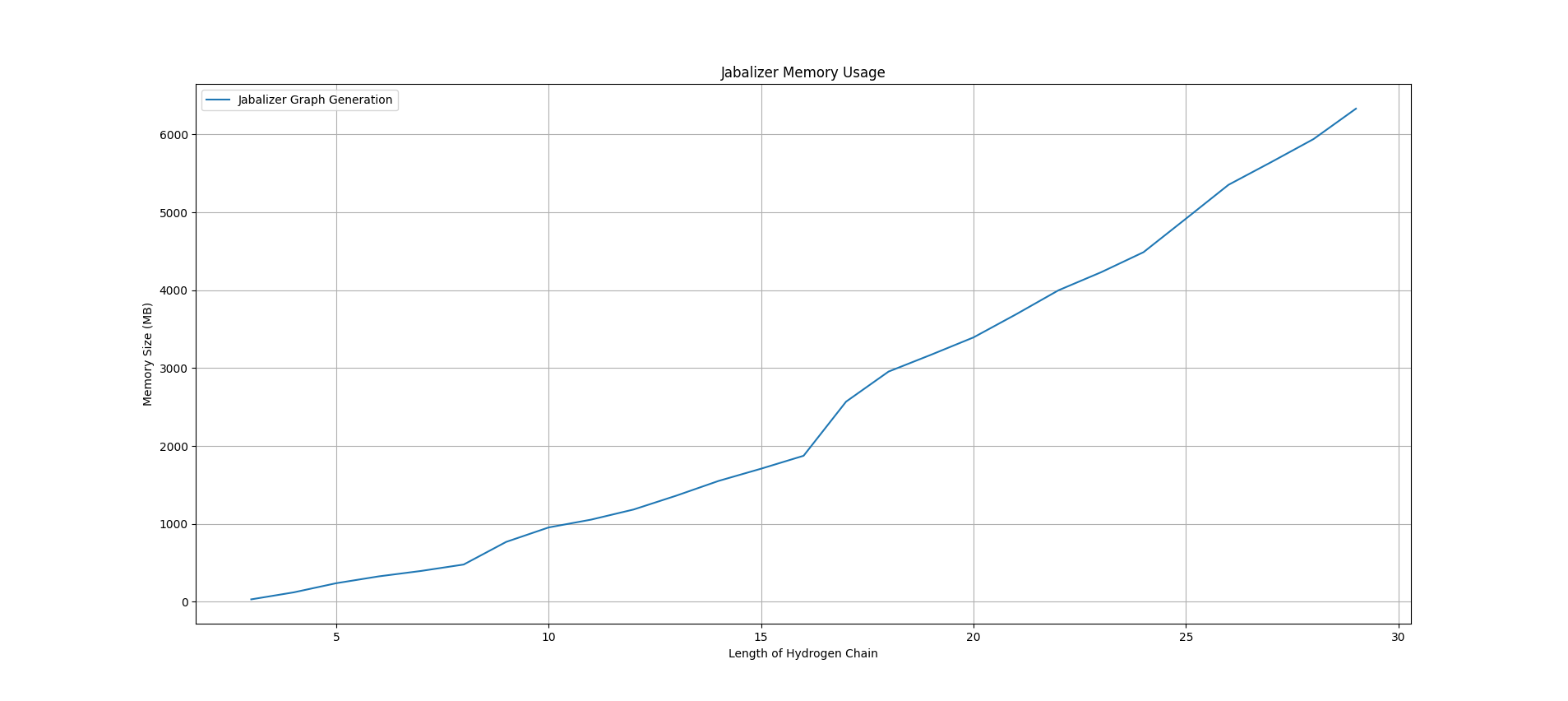}
   \caption{Jabalizer memory use from hydrogen chain simulation.  Test input circuits were size nine to 45 qubits, each with 41-76K gate operations as run on a Macbook Pro 16 with intel i7 processor.}
   \label{fig:jabalizer_memory}
\end{figure*}
\begin{figure*}
   \centering
   \includegraphics[width=\linewidth]{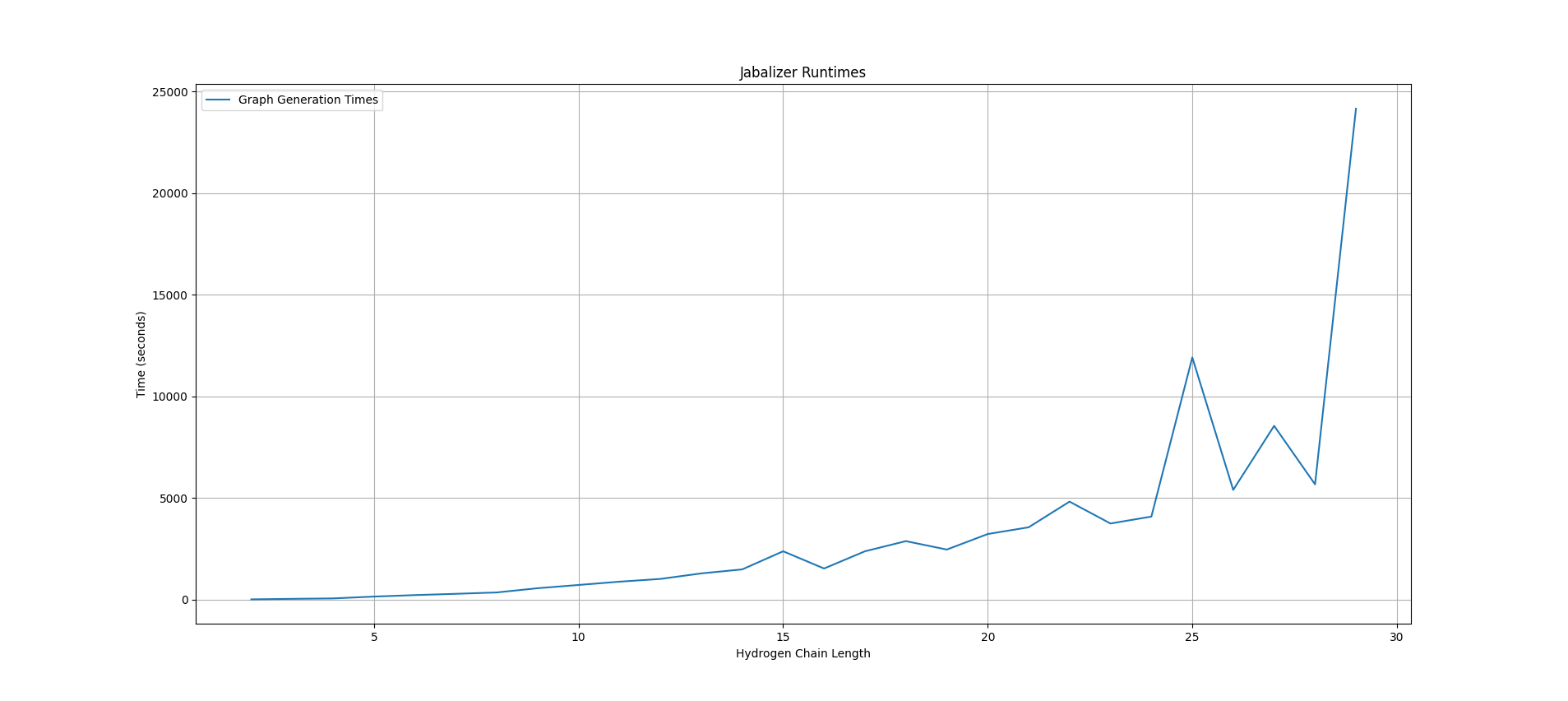}
   \caption{Jabalizer runtimes of hydrogen chain simulation.  Test input circuits were size nine to 45 qubits, each with 41-76K gate operations as run on a Macbook Pro 16 with intel i7 processor.}
   \label{fig:jabalizer_runtime}
\end{figure*}

A full introduction to Jabalizer's functionality is available \cite{vijayanCompilationAlgorithmSpecificGraphStates2022} but as a prec\'{i}s, Jabalizer takes a circuit as input to output a graph state consistent with the fault-tolerant ICM form  \cite{PalerFaultTolerantHighLevelQuantumCircuitsFormCompilationDescription2017}.  Initialisation-CNOT-Measurement (ICM) is a technique of decomposing then measuring an input algorithm/circuit as a requirement of fault-tolerant computation.  Jabalizer realises the ICM form through an algorithm-specific graph state (henceforth, ASG).  To fulfill the ASG strategy, Jabalizer passes the Clifford gates component of its input circuit to a classical computing facility, to resolve using the stabiliser formalism \cite{gottesmanHeisenbergRepresentationQuantum1998,raussendorfFaultTolerantOneWayQuantumComputer2006}; Jabalizer passes all remaining $T$ operators to quantum hardware as a specification for the requisite (magic state) distillation lattice.  The Clifford/non-Clifford outputs are reconciled as the final stage of computation.  Prima facie, reducing a cluster state to ASG with Jabalizer is an efficient compiling solution.  

\begin{figure*}
   \centering
   \includegraphics[width=\linewidth]{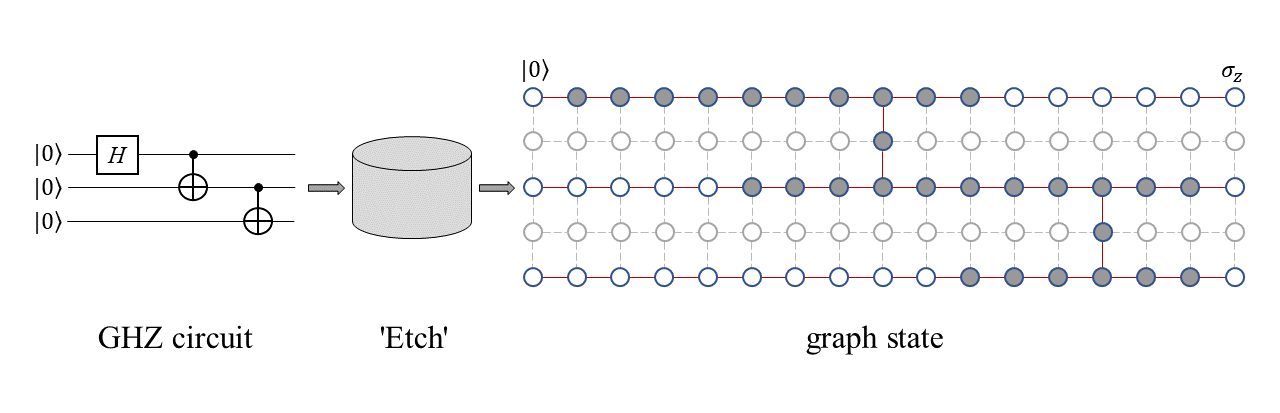}
   \caption{Circuit etching by input and output.  The GHZ circuit is passed through the Etch rules engine to obtain the equivalent graph state.  Qubits (vertices) of the graph state are represented by circles while entanglement with neighbouring qubits is represented by an unbroken line (edge) connecting them.  The leftmost column of the graph state, labelled $\ket{0}$, consists of the input qubits, which are prepaered in the zero state as per the circuit.  The rightmost column of the graph state consists of the readout qubits and is labelled $\sigma_{z}$ to indicate they will be measured in the computational basis.  Qubits with a solid fill represent X or Y measurements while qubits with no fill, excepting the readout qubits, are X measurements.  Note, faint, \enquote{greyed-out} qubits will be removed from the graph state by Z-measurements in advance of measuring computational qubits; any entanglement with computational qubits will be lost upon Z-measurement.}
   \label{fig:Etch_functions}
\end{figure*}

Zapata Computing (Boston, MA) ran tests of Jabalizer as part of an exercise in benchmarking compilation times.  The tests were to simulate a hydrogen chain of length two to 30 through quantum signal processing (QSP) \cite{lowMethodologyResonantEquiangularCompositeQuantumGates2016,lowOptimalHamiltonianSimulationQuantumSignalProcessing2017} and exposed potentially insoluble limitations to global compiling on Jabalizer.  Test input circuits of size nine to 45 qubits, each with 41-76K gate operations, were passed to Jabalizer.  Memory crashes when writing to intermediate representations\footnote{An intermediate representation in classical computing is a data structure to which compilation rules are successively applied.}, or prohibitive runtimes from processing the Clifford gate component resulted from input circuits upwards of 30 qubits in size.  Summary statistics of the tests appear as Figures \ref{fig:jabalizer_memory} and \ref{fig:jabalizer_runtime}, below.  If Jabalizer is to be a plausible solution for any non-trivial quantum computational problem, the application needs to be robust enough to process circuits of much larger dimensions than those administered as part of the QSP tests.

Jabalizer's strength, as a complete implementation of compiling to the ICM form, is also its weakness namely, the application requires that its input qubits be prepared in state $\ket{0}$, which necessarily precludes composition of its output patterns.  Jabalizer's problem with memory crashes or prohibitive runtimes cannot be circumvented through composition (e.g. breaking up a large circuit into smaller sub-circuit components, say sub-circuit\textsubscript{A} through sub-circuit\textsubscript{Z}, passing sub-circuit\textsubscript{A} to the respective classical/graph-state quantum computing facilities, for its output then to serve as input to sub-circuit\textsubscript{B} et cetera).  A solution involving distributed (classical) computing is theoretically plausible but would still necessitate an overhaul of the Jabalizer application.

Once the QSP simulations exposed the problems with Jabalizer, circuit etching was proposed as an alternative compiling solution.  As the name would suggest, circuit etching involves transpiling a circuit into an equivalent $\ket{G}$.  As a basis for comparison, \enquote{equivalence} is a measure of parity between circuit and graph state by computational output, not by speed or resource requirements.  The application, Etch takes a circuit encoded in JSON format as input which it then transplies to $\ket{G}$ output.  When interpreting Etch's output, 
\begin{itemize}
   \item (input) circuit configuration strictly specifies the size of (output) graph state: Etch does not optimise the specification of its graph state layout; and
   \item $\ket{G}$ output complies with the two-dimensional lattice of \cite{raussendorfOneWayQuantumComputer2001,raussendorfComputationalModelUnderlying2002,raussendorfMeasurementbasedQuantumComputation2003} but not the three-dimensional form associated with error-correcting strategies (i.e. \cite{raussendorfFaultTolerantOneWayQuantumComputer2006}); and
   \item Etch would pass $\ket{G}$ in its entirety, both Clifford and non-Clifford, to the quantum computing facility for it to resolve.  In other words, a classical computing facility processes no component of Etch's $\ket{G}$ output.
\end{itemize}
\noindent
Etch is therefore a tool for quantifying circuit etching, specifically the efficiency of circuit etching with regard to qubit resources, as an alternative to ASG.  Figure \ref{fig:Etch_functions} is a representation of Etch functions and the layout of Etch graph state output.

Even though QSP tests have exposed its real limitations, to repeat, Jabalizer's small payload to quantum hardware, composed only of \textit{T} patterns, would still seem more efficient with qubit resources as compared with Etch's monolithic payload.  Let $\lambda$ represent an arbitrary quantum algorithm expressed in the universal quantum gate set and passed to Jabalizer or Etch.  The algorithm $\lambda$ will have a proportion of Clifford patterns, $\lambda_{\mathcal{P}}$ and $T$ patterns, $\lambda_T$ such that $\lambda = \lambda_{\mathcal{P}} + \lambda_T = 1$.  Ignoring 
\[
\lambda = \lambda_T;\hspace{1mm} \lambda = \lambda_{\mathcal{P}},
\]
it seems obvious that processing $\lambda$ (i.e. circuit etching) will always consume more qubits than processing $\lambda_T$ (i.e. algorithm-specific graph).  Circuit etching as an alternative to ASG thus hinges upon the \textit{relative} efficiency of initialising $T$-states and specifically, the total number of qubits required.  Both ASG and circuit-etching require the same number of qubits to process $\lambda_T$.  For its part, Jabalizer processes $\lambda_T$ with \enquote{single-use} qubits because a classical computing facility will process $\lambda_{\mathcal{P}}$.  In comparison, circuit etching is a proposal to recycle any qubit consumed in processing $\lambda_{\mathcal{P}}$ for the purpose of then processing $\lambda_T$.

Once measured, a qubit effectively has no further input to a graph state computation (cf. \cite{browneResourceEfficientLinearOpticalQuantumComputation2005}).  Under circuit etching, that measured qubit might be repurposed as a component of a $T$ factory.  In other words, qubits identified through Etch as part of its output $\ket{G}$ could be recycled to process $T$ operations required by $\ket{G}$.  From this point of view, even a retooled Jabalizer has the hidden cost of the single-use qubits required to process its $T$ operations.  It bears repeating that Etch works with logical qubits.  Temporarily suspending questions of coherence times for a recycled qubit, if the ratio of Pauli to non-Pauli measurements is close enough to an acceptable distillation ratio then, resource allocation under circuit etching becomes multi-functional as distinct from mono-functional under ASG.

The \enquote{concatenation protocol} is an instance of those \textit{higher} fidelity magic state distillation circuits named above.  The protocol takes 15 \textit{logical} qubits, each encoded under 15-to-1 distillation, to apply the same 15-to-1 encoding process to them in turn and thereby realise a probability of error of 35(35\textit{p}\textsuperscript{3})\textsuperscript{3} to leading order.  On paper, a single concatenation protocol would amount to a 225-tiles block (i.e. $15^2$) however Litinski (Figure 19, \cite{litinskiGameSurfaceCodes2019}) proposes a concatenation protocol of a 176-tiles block of dimensions 11 rows by 21 columns.  This 176-tiles block is an instance of a $T$ factory, as identified above.

Consider again the two-dimensional lattice of $\ket{G}$, that can be described as an amalgam of qubits arranged in consecutive columns or, equivalently arranged in rows.  If $\ket{G}$ is the resource for an algorithm $\lambda = \lambda_{\mathcal{P}} + \lambda_T$ then, at least one $T$ pattern is sited within $\ket{G}$ and will require a sufficient number of unused qubits to encode it through a $T$ factory in this case, Litinski's 176-tiles concatenation protocol (cf. \cite{elmanOptimalSchedulingGraphStatesPathDecompositions2024}).  To encode the $T$ pattern's readout qubit will require 11 other unused qubits as a contiguous block, in the same column as it from which might be formed a column of Litinski's 176-tiles concatenation protocol.  The site of the $T$ pattern within the lattice is irrelevant so long as there are sufficient unused qubits to create Litinski's [11, 21] tile block adjacent to it.  In other words, circuit etching must repurpose 11 qubits that were measured in the Pauli basis, for every one non-Clifford readout qubit or be within an arbitrary level of tolerance to this ratio (say, no greater than 25 Pauli : 1 non-Pauli).  \textit{The distillation ratio mooted above is thus 11 : 1} and represents the target of testing outlined in section \ref{evetch}, below.  Furthermore, circuit etching must approximate the 11 : 1 distillation ratio regardless of the proportion of $\lambda_T$ and $\lambda_{\mathcal{P}}$ operators.  Iff circuit etching satisfies both of these conditions, it is an efficient resourcing alternative to ASG. 
\begin{figure*}
   \centering
   \includegraphics[width=\linewidth]{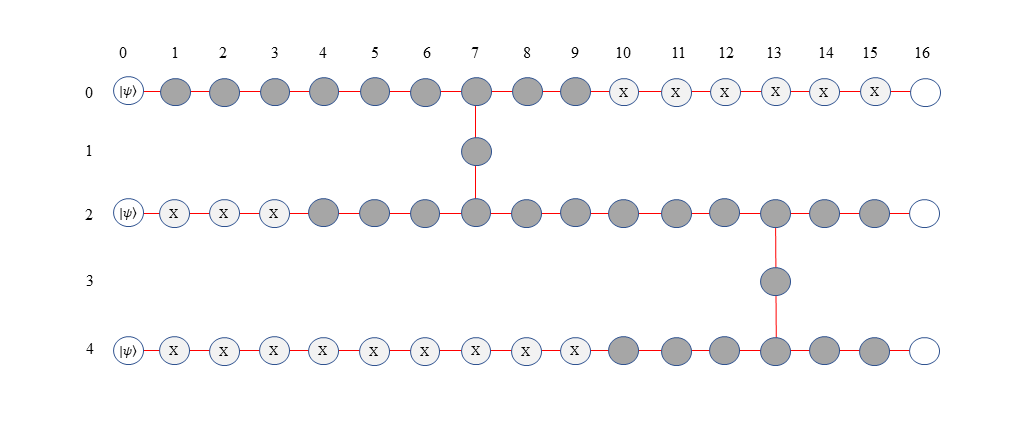}
   \caption{Close up of Etch output $\ket{G}$ of Figure \ref{fig:Etch_functions}: superfluous qubits and entanglement have been removed by measurement in basis $\sigma_z$.  Input qubits (column 0) are prepared in state $\ket{\psi}$; only the $QC_\mathcal{C}$ readout qubits (column 16) are distinct as pattern input-/output qubits are combined.  Qubits with a solid fill represent measurements in basis $\sigma_x$ or $\sigma_y$; all other qubits, excepting the readout qubits, are \enquote{wire} measurements \cite{raussendorfComputationalModelUnderlying2002} in basis $\sigma_x$, to propagate state through $QC_\mathcal{C}$.}
   \label{fig:expand_G}
\end{figure*}

\section{evaluating circuit etching}\label{evetch}
The Etch functions relevant to this paper include,
\begin{enumerate}[label=(\roman*)]
   \item create an equivalent $\ket{G}$ output from a circuit model input.  As noted above, the layout of measurement patterns is set by the input circuit and without spatial optimisation; and
   \item specify the minimum dimensions of $\ket{G}$ necessary to processing (i).
\end{enumerate}

The following gates are valid circuit input:
\begin{itemize}
   \item adjacent CNOTs, exclusively,
   \item $R_x(\theta)$, $R_y(\theta)$, $R_z(\theta)$ as single-qubit rotations by $\theta$ radians about axis $X/Y/Z$, respectively,
   \item Hadamard,
   \item $S/S^{\dagger}$,
   \item \enquote{general rotation}, $U_{Rot}$, 
   \item $T/T^{\dagger}$.
\end{itemize}

Etch lays out the patterns of its graph state output from left-to-right, top-to-bottom.  The property of composition means that horizontally adjacent patterns have their readout and input qubits combined (see Figure \ref{fig:expand_G}); only $QC_\mathcal{C}$ readout qubits are distinct.  Further to the restriction of adjacent CNOTs, \enquote{non-neighbouring} CNOTs pose two problems for the Etch model.  First, the non-neighbouring CNOT blocks the left-to-right teleportation of state of any qubit wire situated between the CNOT control and target qubit wires.  Notionally, this problem is soluble with a crossing sub-pattern proposed in \cite{raussendorfMeasurementbasedQuantumComputation2003} but that solution involves a greater number of qubits than the standard 15-qubits CNOT and therefore also increases the ratio of Pauli to non-Pauli measurements in $QC_\mathcal{C}$.  Second, including non-adjacent CNOTs also increases wastage of qubits because the number of waste qubits (i.e. qubits removed from $QC_\mathcal{C}$ through measurement in basis $\sigma_z$) grows proportionally with lattice size\footnote{ Note, excised $Z$-measurement qubits do not count towards the distillation ratio}.

\begin{table*}
   \centering
   \begin{filecontents*}{patterns_layout/sample_circuits_data.txt}
col1,col2,col3,col4,col5,col6,col7,col8,col9,col10,col11
1,"[8, 76]",608,14,6,4,7,222,386,375,34.1
2,"[38, 348]","13,224",66,32,12,40,"5,536","7,688","7,636",146.8
3,"[22, 204]","4,488",41,18,6,22,"1,818","2,670","2,642",94.4
4,"[161, 1500]","241,500",323,138,39,148,"103,362","138,138","137,951",737.7
5,"[14, 136]","1,904",23,12,4,16,804,"1,100","1,080",54.0
6,"[145, 1232]","178,640",259,116,34,154,"71,340","107,300","107,112",569.7
7,"[154, 1420]","218,680",286,132,39,159,"93,588","125,092","124,894",630.8
8,"[99, 928]","91,872",182,86,26,103,"39,818","52,054","51,925",402.5
9,"[134, 1228]","164,552",256,114,27,135,"69,882","94,670","94,508",583.4
10,"[95, 892]","84,740",166,84,18,111,"37,380","47,360","47,231",366.1
11,"[47, 448]","21,056",76,42,14,55,"9,366","11,690","11,621",168.4
12,"[139, 1280]","177,920",268,118,33,135,"75,402","102,518","102,350",609.2
13,"[37, 376]","13,912",70,34,10,37,"6,358","7,554","7,507",159.7
14,"[60, 524]","31,440",112,48,13,60,"12,528","18,912","18,839",258.1
15,"[105, 976]","102,480",199,90,25,107,"43,830","58,650","58,518",443.3
16,"[121, 1100]","133,100",245,100,33,104,"54,900","78,200","78,063",569.8
17,"[47, 344]","16,168",86,32,8,50,"5,472","10,696","10,638",183.4
18,"[76, 708]","53,808",161,64,16,65,"22,592","31,216","31,135",384.4
19,"[138, 1196]","165,048",255,112,32,140,"66,864","98,184","98,012",569.8
20,"[171, 1540]","263,340",311,144,49,175,"110,736","152,604","152,380",680.3
21,"[63, 544]","34,272",118,50,20,56,"13,550","20,722","20,646",271.7
22,"[117, 1096]","128,232",227,100,32,110,"54,700","73,532","73,390",516.8
23,"[97, 932]","90,404",180,86,25,98,"39,990","50,414","50,291",408.9
24,"[109, 1048]","114,232",211,96,33,99,"50,208","64,024","63,892",484.0
25,"[36, 324]","11,664",56,30,9,43,"4,830","6,834","6,782",130.4
26,"[18, 156]","2,808",28,14,6,21,"1,078","1,730","1,703",63.1
27,"[87, 836]","72,732",176,76,20,81,"31,692","41,040","40,939",405.3
28,"[30, 268]","8,040",54,24,10,27,"3,192","4,848","4,811",130.0
29,"[79, 760]","60,040",150,70,12,84,"26,530","33,510","33,414",348.1
30,"[128, 1188]","152,064",220,112,30,148,"66,416","85,648","85,470",480.2
\end{filecontents*}
\pgfplotstabletypeset[
   1000 sep={,}
   ,col sep=comma
   ,every head row/.style={
   after row={
   circuit & & & & &arbitrary & &excised &graph & &Pauli : \\
   id &specification &lattice &Clifford$_4$ &CNOT$_{6-1-6}$ &Z-rotation$_4$ &T$_4$/T\textsuperscript{\textdagger}$_4$ & Z-measurements &state &Pauli &non-Pauli \\
   \midrule}
   ,before row={\toprule}
   ,column type={m{6em}}
   ,output empty row
   }
   ,every last row/.style={
   after row=\bottomrule
   }
   ,text indicator=",string type
   ]{patterns_layout/sample_circuits_data.txt}
   \caption{Etch-generated layouts of graph states from IQP circuits input.  Details of each column are found in the body of the paper.  Columns \textbf{lattice}, \textbf{Clifford\textsubscript{4}}, \textbf{CNOT\textsubscript{6-1-6}}, \textbf{arbitrary Z-rotation\textsubscript{4}}, \textbf{T/T\textsuperscript{\textdagger}\textsubscript{4}}, \textbf{excised Z-measurements} are raw counts by property, subscripts denote the number of qubits in a pattern;
   columns \textbf{graph state} and \textbf{Pauli} are derived values.  Each instance of input IQP circuit features every (IQP circuit) gate but in different proportions and therefore the same \textit{combinations} of patterns appear in each graph state output.  The distillation ratio of each graph state appears as column \textbf{Pauli : non-Pauli}.}
   \label{tab:table1}
\end{table*}

Input circuits passed to Etch as part of its distillation ratio evaluation were modelled on the instantaneous quantum polynomial (IQP) circuit \cite{shepherdTemporallyUnstructuredQuantumComputation2009}.  The qubits count of each circuit was generated at random\footnote{the Python 3.9.2 random library, used to set each circuit's qubits size, is actually a pseudo-random number generator.} from within a range, 5-120 (cf. \cite{crossValidatingQuantumComputersUsingRandomizedModelCircuits2019}).  Note, the combinations of circuit gates is consistent, regardless of randomness in circuit sizes: the same gates, albeit in different proportions, appear in each IQP circuit instance.  A circuit entirely generated at random is unrepresentative of a problem-defined algorithm and indeed, algorithms working in the same hypothetical problem space are commonly consistent in combinations of gates. 

Thirty instances of IQP circuit were transpiled through Etch to test the capacity of circuit-etching to achieve the distillation ratio.  Data of the graph state output, including breakdown by gates count and pauli to non-pauli ratios appear as Table \ref{tab:table1}.  Columns of Table 1 are to be interpreted as follows,

\begin{itemize}
   \item \textbf{circuit id}, unique id of the IQP test circuit;
   \item \textbf{specification}, dimensions by [row, column] of Etch graph state output.  The reader is reminded that Etch specifies its graph state output as a two-dimensional lattice and strictly in accordance with layout of the input circuit.  In other words, lay out of Etch's graph state output is not optimised;
   \item \textbf{lattice}, the total number of qubits in Etch's graph state output, as derived from \textbf{specification}, prior to superfluous qubits being removed by Z-measurement (see Figure \ref{fig:Etch_functions});
   \item \textbf{Clifford\textsubscript{4}}, \textbf{CNOT\textsubscript{6-1-6}}, \textbf{arbitrary Z-rotation\textsubscript{4}}, \textbf{T/T\textsuperscript{\textdagger}\textsubscript{4}}, the count of this pattern type to appear in the graph state that is derived from IQP circuit input.  The subscript to each column heading denotes the number of qubits in a pattern.  Qubits of a pattern appear consecutively in a row hence the separation of \textbf{Clifford\textsubscript{4}} and \textbf{CNOT\textsubscript{6-1-6}}: a CNOT pattern spreads over three rows of 6 then 1 then 6 (see Figure \ref{fig:expand_G} for examples of this CNOT configuration).  \textbf{T/T\textsuperscript{\textdagger}\textsubscript{4}} is separated from \textbf{arbitrary Z-rotation\textsubscript{4}} on the grounds that the former is a specific type of the latter; 
   \item \textbf{excised Z-measurements}, the number of qubits that are superfluous to the graph state and to be removed by measurement in basis $\sigma_z$.  These qubits are counted in \textbf{lattice};
   \item \textbf{graph state}, the number of the computational qubits in Etch's graph state output, derived as \textbf{lattice} \textit{less} \textbf{excised Z-measurements};
   \item \textbf{Pauli}, the number of Clifford operation qubits in \textbf{graph state}, derived as \textbf{graph state} \textit{less} (\textbf{arbitrary Z-rotation\textsubscript{4}}  \textit{plus} \textbf{T\textsubscript{4}/T\textsuperscript{\textdagger}\textsubscript{4}});
   \item \textbf{Pauli : non-Pauli}, ratio of Pauli to non-Pauli measurements: \textbf{pauli} over (\textbf{arbitrary z-rotation\textsubscript{4}}  \textit{plus} \textbf{T\textsubscript{4}/T\textsuperscript{\textdagger}\textsubscript{4}}).  This value is the distillation ratio.
\end{itemize}

The 30 circuits of Table \ref{tab:table1} are a reliable sample for the purposes of extrapolating to the population \cite{gujaratiEssentialsEconometrics1999}.  With $\bar{X}$ denoting the sample mean, the following statistics can be obtained for column heading: 
\begin{itemize}
   \item $\bar{X}$, \textbf{graph state}: 50,966.1
   \item $\bar{X}$, \textbf{Pauli}: 50,858.5
   \item $\bar{X}$, \textbf{Pauli : non-Pauli}: 362.8
\end{itemize}
\noindent
None of the 30 input circuits satisfies or approximates the distillation ratio of 11:1.  The average \textbf{graph state}, comprising almost 51,000 vertices (qubits) is approximately 97-99 rows by 928-932 columns, depending upon the combination of input gates, with \textbf{Pauli : non-Pauli} over 360, far in excess of the desired 11:1 ratio.  Indeed, assuming the population is normally distributed and $\bar{X}$ \textbf{Pauli : non-Pauli} and its standard deviation (209.0) are reliable estimators of it, the probability of Etch producing the distillation ratio according to the above circuit input parameters amounts to 4.63E-04.

\section{discussion}\label{Disc}
By the \textbf{Pauli : non-Pauli} data of Table \ref{tab:table1}, circuit etching is an inefficient strategy for processing a given $\ket{G}$, at least as compared with ASG.  The gap between the target distillation ratio and observed \textbf{Pauli : non-Pauli} values is invariably large and the data resolve to a minuscule probability that circuit etching could scale with or even attain the target distillation ratio.  In brief, nearly all cases of circuit etching captured as Table \ref{tab:table1} produce more qubits eligible for recycling than would ever be required.  Indeed, one could \textit{halve} each of the \textbf{Pauli} values reported in Table \ref{tab:table1} to then derive an adjusted \textbf{Pauli : non-Pauli} value and still none of the graph states would return the desired ratio\footnote{halving their pauli value sets two (2) circuits - ids 1 and 5 - at (adjusted) Pauli : non-Pauli ratios of 17:1 and 27:1, respectively.}.  Moreover, for any future improvement to magic state distillation techniques that lowers the number of qubits required (e.g. \cite{litinskiMagicStateDistillation2019}), circuit etching is only made more inefficient as a strategy.  In the interest of completeness, it is worth exploring whether optimising Etch's layout patterns might make the application more competitive with Jabalizer.  

As stated above, Etch does not optimise the specification of its graph state layout but strictly adheres to the configuration of an input circuit.  It is worth noting that many, if not all, of the patterns laid out by Etch could be spatially or compositionally optimised.  It was a deliberate decision to not manually optimise Etch's output patterns before lattice specification and resultant qubit counts: any optimisation put in place, or combination thereof, may not have been the best optimisation (cf. \cite{hietalaVerifiedOptimizerForQuantumCircuits2021}), which might then open Etch to \enquote{if only} criticisms.  Equally, however, it is prudent to estimate the benefits of \enquote{what if} criticisms.  For example, efforts at reducing the size of the lattice (i.e. \textbf{specification} of Table \ref{tab:table1}) would do nothing to affect the \textbf{Pauli : non-Pauli} value because it would not reduce the absolute count of patterns, which actually determines the ratio's numerator.  Only reducing the number of Pauli qubits will optimise Etch's $\ket{G}$ output with respect to the distillation ratio and conceivably, marginally increase space efficiency of the lattice.  Taking a different tack, the Pauli numerator could be reduced by substituting patterns, the most obvious being switching the 13 qubits CNOT pattern for the eight qubits CZ pattern \cite{raussendorfMeasurementbasedQuantumComputation2003}.  Again the answer must be a qualified \enquote{yes} because CNOT is not the commonest pattern in Table \ref{tab:table1}.  Only five circuit instances have CNOT counts which tally within 50-55 per cent of counts of \textbf{Clifford\textsubscript{4}} pattern in the same circuit: ids 5, 10, 11, 25 and 30.  Every other circuit has CNOT pattern tallies that are fewer than half the number of \textbf{Clifford\textsubscript{4}} patterns.  By qubits a CNOT pattern does contribute more to the \textbf{Pauli} count than a CZ pattern but in absolute number, the CNOT pattern (and by proxy, the CZ pattern) was not the biggest contributor to \textbf{Pauli}.

Of course, it could be argued that surpassing the distillation ratio means circuit etching has extra capacity for $T$-factories which could be put to use as a temporal optimisation of $\ket{G}$.  The tests discussed in this paper did not explore a scenario of extra capacity although it is worth observing that having extra capacity and needing it do not always coincide.  It would seem that optimisation alone is not the key to a more competitive circuit etching strategy.

Before considering the options left for graph-state computation strategies, it must be stressed that real limitations to such strategies do not repudiate the graph-state computation model any more than they implicitly validate the circuit model.  The tests of Jabalizer and Etch outlined above only show that neither application in its current format brings an inexpensive route to resourcing $T$ operations.

A compilation strategy that combines aspects of the ASG and circuit-etching strategies may work for graph-state computing.  A hybrid compilation strategy is predicated upon,
\begin{enumerate}
   \item ASG not being bound to particular patterns of its Clifford or non-Clifford operators; and
   \item Circuit etching's property of composition of its patterns.
\end{enumerate}
From this viewpoint, Jabalizer could be reconfigured as a \textit{pattern factory}.  By pattern factory is meant (Jabalizer) producing graph states, within its compile limitations, that are components of a computation.  The closest analogue to what is meant by a \enquote{component} of computation is the sub-routine of classical computing.  In other words, Jabalizer's pattern\textbackslash sub-routine need not be confined to replicating a single gate operation so long as it does not transgress identified compilation limits.  Just as patterns can be composed, in this case into components, Jabalizer's pattern\textbackslash sub-routine can also be decomposed when memory use or runtimes become prohibitive.  

As detailed above, Etch works as a tool for testing the circuit etching strategy and especially, the viability of qubit recycling.  Furthermore, it works within strict parameters of the two-dimensional lattice and Clifford patterns prescribed by \cite{raussendorfMeasurementbasedQuantumComputation2003}.  Neither of these restrictions is necessary to a circuit etching strategy.  The next version of Etch could be re-scoped to be a scheduling tool that,
\begin{enumerate}[label=(\roman*)]
   \item composes Jabalizer's pattern output(s) to form $\ket{G}$; then
   \item passes (i) or streams sub-graphs of it to the quantum computing facility for processing.
\end{enumerate}
Further to (i), a pattern need not be complete and self-contained, there just needs to be correct preparation of the input qubit(s) of the graph being passed to the quantum computing facility at a given point in time.  Indeed, the streaming of (ii) is consistent with MBQC principles so long as the the order and basis of measurement within and between sub-graphs does not vary from that necessary to (i).  It is the composability of patterns that enables streaming of sub-graphs to work.

This hybrid strategy bypasses the problem of a monolithic $\ket{G}$ and the need for either ASG or qubit recycling to mitigate it.  Rules to parallel process $\ket{G}$ as an optimisation of time efficiency could be introduced (e.g. \cite{evansMCBethMeasurementBasedQuantumProgrammingLanguage2023}) although Etch, as a receiver of patterns, would still have no capacity to optimise for space\textbackslash resources.  Finally, whether (ii) passes synchronously or asynchronously to it would depend upon the engineering of the target quantum computing facility.

\begin{acknowledgments}
The views, opinions and/or findings expressed are those of the author(s) and should not be interpreted as representing the official views or policies of the Department of Defense or the U.S. Government.  This research was developed with funding from the Defense Advanced Research Projects Agency [under the Quantum Benchmarking (QB) program under award no. HR00112230007 and HR001121S0026 contracts].
\end{acknowledgments}

\nocite{*}

\bibliographystyle{apsrev4-2}
\bibliography{refs}

\end{document}